# Optically driven rotation of exciton-polariton condensates

Yago del Valle Inclan Redondo[1,2], Christian Schneider[3], Sebastian Klembt[4], Sven Höfling[4], Seigo Tarucha[1], and Michael D. Fraser [1,2][†]

[1] *RIKEN Center for Emergent Matter Science, Wako-shi, Saitama 351-0198, Japan*

[2] *Physics & Informatics Laboratories (PHI Lab), NTT Research, Inc., Sunnyvale, CA 94085, USA*

[3] *Institute of Physics, University of Oldenburg, D-26129 Oldenburg, Germany*

[4] *Technische Physik, Physikalisches Institut and Wilhelm Conrad Roentgen-Research Center for Complex Material System, Universität Würzburg, Am Hubland, D-97074 Würzburg, Germany*

[†] Corresponding author: michael.fraser@riken.jp

## Abstract

The rotational response of quantum condensed fluids is strikingly distinct from rotating classical fluids, especially notable for the excitation and ordering of quantized vortex ensembles. Although widely studied in conservative systems, the dynamics of rotating open-dissipative superfluids such as exciton-polariton condensates remain largely unexplored, as it requires high-frequency rotation whilst avoiding resonantly driving the condensate. We create a rotating polariton condensate at GHz frequencies by off-resonantly pumping with a rotating optical stirrer composed of the time-dependent interference of two frequency-offset, structured laser modes. Acquisition of angular momentum exceeding the critical $1\ \hbar$/particle is directly measured, accompanied by the deterministic nucleation and capture of quantized vortices with a handedness controlled by the pump rotation direction. The demonstration of controlled optical rotation of a spontaneously formed polariton condensate enables new opportunities for the study of open-dissipative superfluidity, ordering of non-Hermitian quantized vortex matter, and topological states in a highly non-linear, photonic platform.

## Introduction

A defining characteristic of superfluid phases of matter (*1*) is their unique response to external forces. Subjected to a longitudinal force, the superfluid flow is frictionless below a critical velocity, and under a transverse force, superfluids are irrotational leading to angular momentum being incorporated as phase defects (quantized vortices) above a critical angular momentum (Hess-Fairbank effect) (*2*). These behaviors are analogous to the phenomena of vanishing resistivity and the expulsion of magnetic fields (Meissner effect) in superconductors, respectively (*3*). Experimentally, superfluid transport and rotation in particle-conserving platforms has been extensively explored in liquid Helium (*4*) and dilute-gas atomic Bose-Einstein condensates (BEC) (*5–7*). In both cases, rotation is typically induced mechanically, either through rotation of the container in the case of He, or via a stirring laser beam in the case of BEC, with the hallmark in both experiments being the incorporation and self-ordering of quantized vortices with increasing rotation speeds.

Semiconductor microcavity exciton-polaritons, hybrid light-matter quasi-particles, are able to form a Bose-Einstein condensate-like state (*8*, *9*) shown to behave as a superfluid (*10*, *11*) and to incorporate vortices in their steady-state condensates, either spontaneously from spatial disorder (*12–14*) or open dissipative effects (*15*, *16*), in vortex-antivortex lattices with zero total angular momentum (*17–20*) or nucleated from an initially coherent polariton injection (*21–24*). Continuous injection of angular momentum has been demonstrated for resonant pumping (*24*), however, pinning effects of the condensate phase to that of the pump laser (*25*) in this geometry make it difficult to completely isolate the superfluid response from the phase coherence of the pump laser. For the case of nonresonant pumping, continuous injection of angular momentum has not been demonstrated but vortices of controllable handedness have been created through the handedness of the external pump (*26*, *27*), an additional pulsed perturbation (*28*), or through non-Hermitian selection of chiral modes (*29*, *30*). A very recent work (*31*) using a rotating optical "bucket" trap has demonstrated that rotation can bias the stochastic formation of *trapped p-modes* into a single vortex mode (of angular momentum $1\hbar/$ particle) with controlled handedness. While this work rotates continuously and off-resonantly, the trap geometry enforces single mode selection and is thus limited to the controlled generation of a single quantized vortex.

Continuous driven rotation of a polariton condensate with spontaneous phase coherence in a geometry permitting free vortex evolution is thus highly sought after as it would allow the nucleation of vortex ensembles enabling the study of nonlinear pattern formation in open-dissipative vortex matter (*7*, *32*), dissipative quantum turbulence (*33*, *34*), the construction of artificial gauge fields (*35*, *36*) and could provide control of recently proposed ring-shaped polariton condensate qubits (*37*). The rapid rotation of a Bose gas when harmonically confined in two-dimensions also mimics a synthetic magnetic field and induces Landau level-like energy spectrum (*7*). In this system, when interactions dominate over other energy scales, a quantum phase transition to a fractional quantum Hall effect (FQH) is predicted (*38*). While studied in atomic gases (*7*, *39*, *40*) and the twisted photonic cavity-Rydberg atom platform (*41*, *42*), achieving the many-body bosonic FQH state remains extremely challenging (*43*). Being intrinsically two-dimensional, combined with its distinct hydrodynamic parameter space (see Supplementary 8) and flexible trapping environment (*44*), the development of techniques to rapidly rotate the polariton condensate may provide an alternative approach to creating a bosonic FQH at elevated temperatures.

A first step towards these goals is the demonstration of controlled quantized vortex nucleation driven by external rotation in analogy to previous experiments in superfluid helium and atomic BEC. A rough estimate for the critical rotation frequency for a polariton condensate can be found by considering the analytical results for a rotating conservative condensate (*45*) and substituting in the polariton hydrodynamic parameters. The range of masses ($\sim 10^{-5} m_e$), nonlinearities and densities in typical experiments suggest that the polariton condensate needs to be rotated at very high frequencies ($\sim 10^9$ Hz, see Supplementary 12) to generate sufficient angular momentum for vortex nucleation. The experimental difficulty of achieving such high frequencies using an incoherent stirring mechanism have thus limited progress.

In this work we used two lasers (see Methods) to create a superposition (Fig. 1A) of a Gaussian (Fig. 1C) and a Laguerre-Gaussian (LG) of orbital charge $l$ (Fig. 1B) (*46*). An orbital charge of $l = 2$ (azimuthal phase winding of $4\pi$) was used with the resulting interference intensity pattern being a ring of $l = 2$ holes that break axial symmetry (Fig. 1D). The addition of a controllable frequency-offset $\Delta f$ causes this intensity pattern to rotate at a fixed rate (*47, 48*), as evidenced by measurements of static frames showing a controlled angle of these holes (Fig. 1D). The frequency offset was controlled by manually tuning the wavelength of one of the lasers within its mode hop-free tuning range, while monitoring $\Delta f$ via the beat signal of the two beams. The superposition beam was used to off-resonantly inject a GaAs microcavity (see Methods). The spontaneously formed polariton condensate is simultaneously pumped and rotated by this "stirring beam". Crucially, this stirring beam does not create an optical trap (in contrast with (*31*)), thus allowing for unconfined movement of vortices and nucleation of vortex ensembles. Two methods were used to quantify the amount of stirring transferred to the condensate. The first was mapping the condensate momentum flow, which provides an ensemble average measurement of angular momentum. The second was phase interferometry, demonstrating the nucleation and capture of quantized vortices. Both measurements demonstrate continuous incorporation of angular momentum exceeding $1\hbar$/particle into a spontaneously formed polariton condensate.

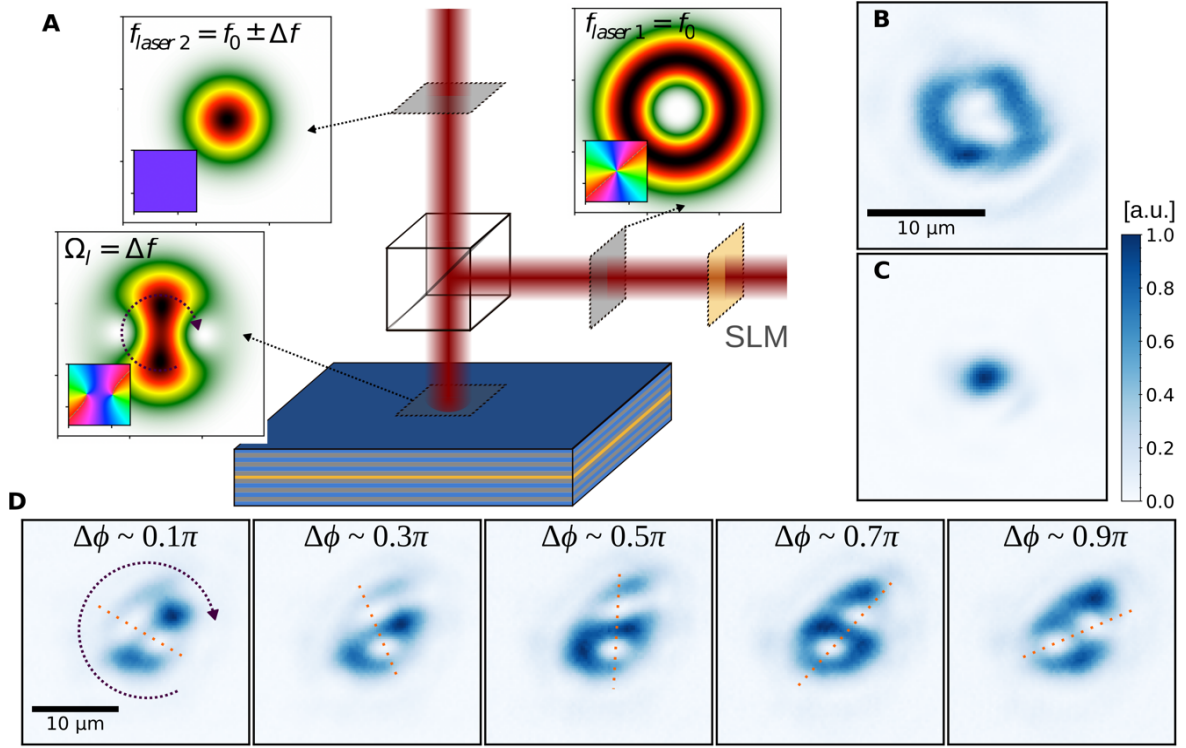

**Figure 1: A rotating pump** is formed by the coherent superposition of a Gaussian and an $l = 2$ Laguerre-Gauss mode with a frequency offset of $\Delta f$. **A** Simplified schematic of pumping process with numerically calculated intensity and phase (inset) of each pump laser and their superposition. Experimental measurements of the **B** LG and **C** Gaussian modes on the microcavity sample surface and **D** a series of temporal snapshots of the experimental rotating superposition profile. The estimated relative phase determined by the angle of the defects is indicated.

# Results

A confocal microscopy setup (see Methods) is used to map the direction and magnitude of condensate flow at all points in space, allowing the detection of azimuthal flow as the hallmark of rotation (Fig. 2A). The microcavity preserves the condensate in-plane momentum ($k_\parallel$) in the emission angle of photoluminescence, permitting the characterization of real and momentum space distributions simultaneously (*49*). A pinhole aperture filters emission at a given spatial location (Fig. 2B), which is then imaged in momentum space. The centroid of the resulting cloud corresponds to the average momentum of polaritons at that position. As expected for a pinhole at $x > 0$, the cloud centroid is shifted to $k_x > 0$ (Fig. 2C) from the radial expansion (*50, 51*) that arises from nonlinear interaction with the nonresonant pump.

As the spread of the condensate cloud in momentum space is comparable to the measured displacements, and further, to distinguish azimuthal flow specifically driven by the rotating pump from any azimuthal flow arising from sample disorder, we carry out a differential analysis between momentum clouds of condensates rotating at the same frequency but with opposite handedness. At low frequency ($|\Delta f| = 20$ MHz, Fig. 2D), the deflection of the condensate momentum cloud is small (~1%) and has a spatial structure that varies with sample position. At high frequency ($|\Delta f| = 3$ GHz, Fig. 2C,E), a transverse shift of the momentum cloud is observed, appearing as two distinct lobes in the map of differential density which are more robust to movements in sample position. This demonstrates that the rotating pump overcomes sample disorder and creates transverse flow in the condensate. Spatially scanning the position of the pinhole shows that the differential flow at high frequency is azimuthal (Fig. 2F) confirming that the entire condensate is experiencing rotation and incorporating angular momentum from the nonresonant pump.

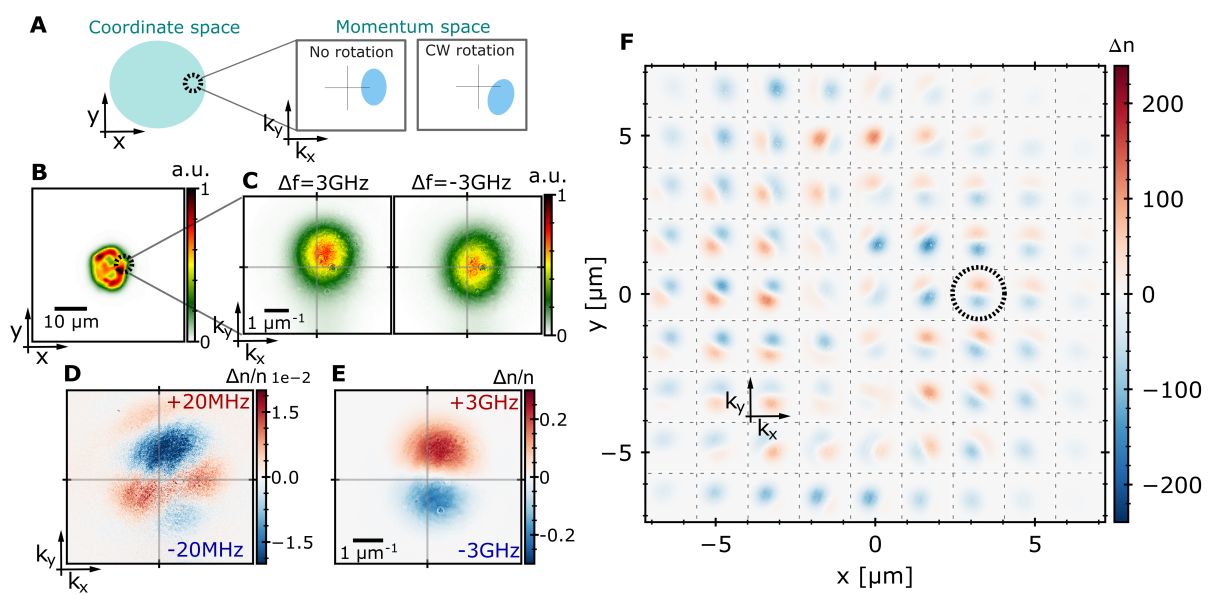

***Figure 2: Differential measurement of condensate rotation. A** Schematic of expected momentum-space clouds for spatially filtered, non-rotating and rotating condensates. **B** Real-space condensate density distribution and **C** measured momentum-space clouds of spatially filtered (black dashed circle) condensate emission for $\Delta f = \pm 3$ GHz. **D** Differential density map at $\Delta f = \pm 20$ MHz and **E** at $\Delta f = \pm 3$ GHz (from data in panel C). **F** Combined differential density maps for a spatially scanned aperture. Measurement in E is circled, with equivalent axis scaling.*

Condensate rotation and its dependency on frequency is further quantified by constructing vector flow maps and estimating their angular momentum (see Methods). Momentum flow maps (Fig. 3A,B) show that the overall handedness of azimuthal flow is controlled by the direction of pump rotation, which can be clearly seen to flip in the spatial maps of angular momentum (Fig. 3D,E). While disorder-induced spatial inhomogeneities exist in these measurements, the change of angular momentum across the measurement area is confirmed to all be in a consistent direction driven by the rotating pump. The total condensate angular momentum per particle ($l = L/N$) also has a handedness consistent with that of the pump at all values of $\Delta f$ (Fig. 3C) and reaches values larger than $1\,\hbar$/particle at the highest frequencies. Notably, the angular momentum plateaus at large $|\Delta f|$, and has an overall response that is slightly asymmetric with respect to the pump rotation, so that for the same frequency, clockwise rotation is more efficient than counterclockwise rotation. Saturation emerges as a dynamical competition between the rotation and the constant divergence of polaritons that is driven by nonlinearity, while the asymmetry has its origin in sample disorder and small misalignment between the two lasers. Both these observations are supported by numerical simulations later in the manuscript.

Single-shot interferometry imaging (see Methods) reveals that the non-resonant rotation was sufficient to nucleate a vortex of charge $\pm 1$, which appears as a forked dislocation in the fringe pattern (Fig. 3G-H). The vortex location and handedness are nominally the same for every realization (see Supplementary 5). For some of the studied frequencies, the phase dislocations exhibit a blurring of the interference visibility in the vicinity of the fork (Fig. 3H), which is exemplary of the time-integrated interferometry of a mobile vortex (*14*). At low frequency, multiple vortices and antivortices are present and they are positioned further away from the condensate center (Fig. 3I). These spontaneously formed vortices have been reported before (*12*) and are pinned by the disordered energy landscape of the microcavity. As the frequency of the rotation is increased, these spontaneous vortices give way to either antivortices ($\Delta f < 0$, Fig. 3G) or vortices ($\Delta f > 0$, Fig. 3H) appearing near the condensate's rotation axis. Plotting the distance of all observed vortices and antivortices from the rotation axis (Fig. 3F) demonstrates how the rotation biases the location and number of vortices, with positive (negative) frequencies preferentially capturing vortices (antivortices) progressively closer to the rotation axis as the rotation speed increases. For some of the negative frequencies ($\Delta f = [-3, -4]$ GHz), more than one co-handed vortex appears near (< 2 μm) the condensate center, which correlates with and can explain why the angular momentum exceeds $1\hbar$/particle at large negative frequencies. The qualitative agreement between the "vortex capture" (Fig. 3F) and the measured angular momentum (Fig. 3C), also provides independent verification that our momentum mapping technique is accurately quantifying the angular momentum of the polariton condensate.

This transition from disorder-nucleated vortices at low frequency to rotation-nucleated vortices at high frequency provides an intuitive explanation for why the angular momentum measured in Fig. 3C is a continuous function of frequency. Vortices away from the rotation axis around which the angular momentum is measured contribute less angular momentum than on-axis vortices and hence, while the intrinsic angular momentum (*52*) of the vortex is still quantized, the total angular momentum around the condensate rotational axis is not. This phenomena is well understood in atomic BEC, where analytical solutions can be extracted for the contribution of off-axis vortices in certain geometries (*45*, *53*) (see Supplementary 10).

The gain-dissipative nature of the polariton condensate allows for the introduction of two further effects that could explain a non-discretized angular momentum: dynamic vortices and transient vortices. Dynamic vortices are those which follow a limit cycle trajectory in the steady state, and lead to the observed extended forks as in Fig. 3H. Since their distance to the origin is not fixed, their contribution to the total angular momentum varies throughout their limit cycle. Transient vortices on the other hand are spontaneously formed by shot noise and are not part of the steady state. They can lead to both positive and negative contributions to the total angular momentum in time averaged experiments depending on their topological charge (see Supplementary 10). Other contributions to angular momentum found in conservative condensates, such as center-of-mass motion and surface wave excitations (*45*) might also be present, but quantifying their influence would require time-resolved experiments.

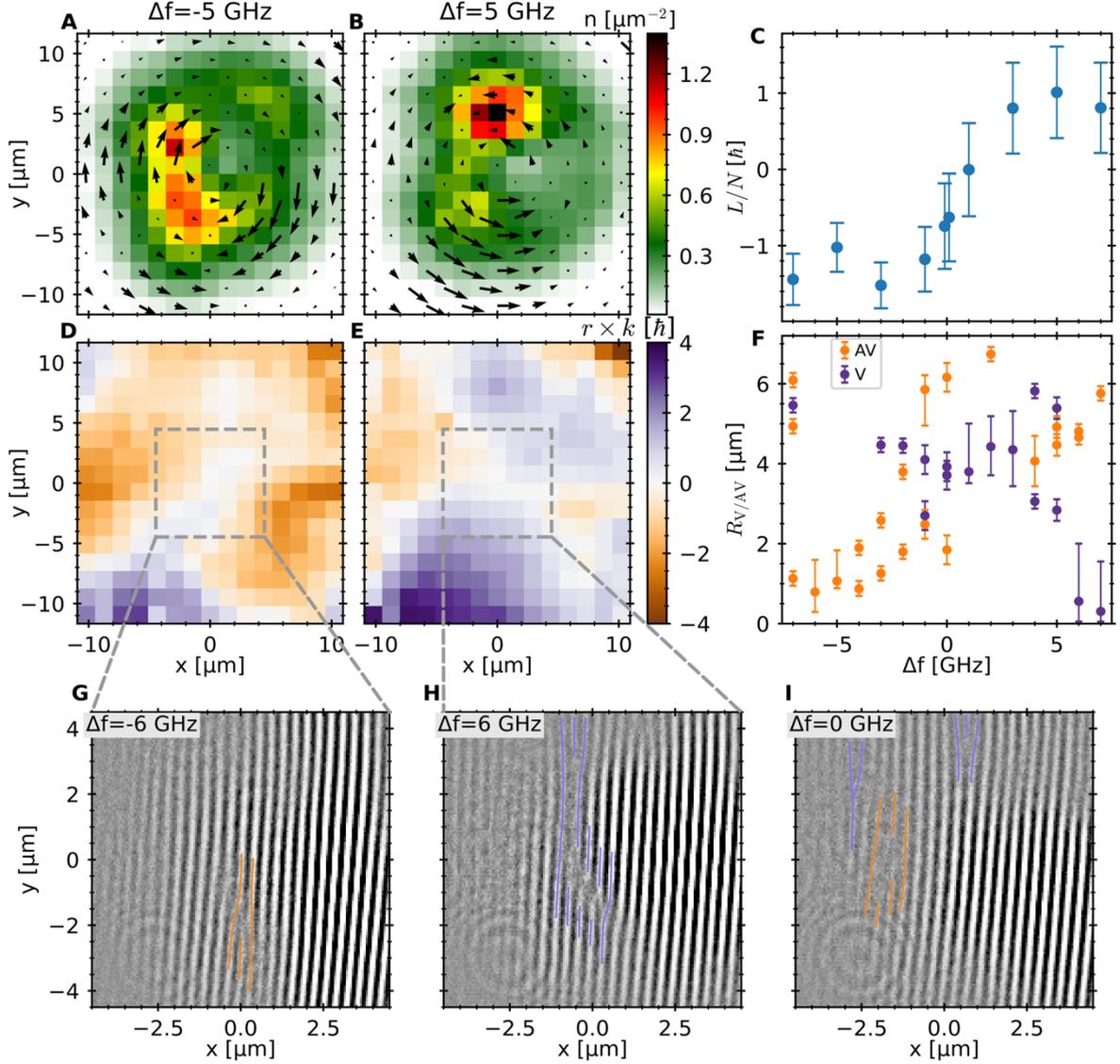

*Figure 3: Frequency dependence of condensate rotation. (A,B) Condensate density and azimuthal velocity map extracted from spatially pin-holed data at two different rotation frequencies. (C) Angular momentum per particle $l = L/N$ across the measurement area as a function of pump rotation frequency. Main source of error is the position of the axis of rotation, which is estimated to be $\pm 1.5$ μm. (D,E) Angular momentum distribution of the velocity maps in (A,B). (F) Vortex positions extracted from interferograms as a function of frequency, distinguished as vortices (V) or antivortices (AV). Error bars are estimated from the maximum of the following: one fringe width, the shot-to-shot variation in vortex core position, and the spatial extent of the blurred fork. (G,H,I) Single-shot interferograms at three different rotation frequencies. Purple (orange) lines are guides to the eye for locating vortices (antivortices).*

To determine the mechanism of optical rotation of the polariton condensate, and to explain the values of angular momentum observed, numerical solutions of an open-dissipative Gross-Pitaevskii model (*54*) (see Methods) are compared with experimental results. Using a pump profile and polariton properties applicable to the experiment, condensate rotation is found over a wide range of parameters. To confirm the experimental assumption that sample disorder leads to asymmetry in the condensate response to rotation with reversed pump rotation direction, the numerical simulations in Figure 4 incorporated a shallow potential (see Methods, Supplementary 9) to emulate realistic microcavity disorder (*55*). For a rotation speed of $\Delta f = \pm 8$ GHz, simulations show that the time-integrated condensate density and vector flow map (Fig. 4A,B), and spatial distribution of transverse flow magnitude (Fig. 4D,E) are qualitatively similar to those in experiment (Fig. 3A,B,D,E). The handedness of the rotation is transferred to the condensate and there are non-axially symmetric spatial inhomogeneities in both density and flow. The accumulation of angular momentum with increasing pump rotation velocity (Fig. 4C), also has a close qualitative match to the experimental data in Figures 3C, exhibiting an asymmetry with rotation direction (blue line in Fig. 4E, $l(+10 \text{ GHz}) = 0.6\hbar$, $l(-10 \text{ GHz}) = -0.5\hbar$). Finally, a brief saturation of $l$ at around $\Delta f \sim 12$ GHz is observed, after which the simulations fail to reach a steady-state rotating condensate.

While the total angular momentum is less than 1 $\hbar$/particle at high frequency (Fig. 4C), the temporal snapshots of the condensate phase show that a vortex is still nucleated, captured and has the same handedness as the pump rotation (Fig. 4G,H). This initially counterintuitive fact can be explained, as before, by the fact that the captured vortex is not on-axis which, in addition to the presence of dynamic, oppositely signed vortices at the condensate periphery, is consistent with the experiments presented in Figures 3G,H. Also consistent with the reduced fringe visibility in the experiments (Fig. 3H), the simulated vortex is dynamic and moves periodically along a stable limit-cycle (gray line in Fig. 4G,H), determined by the disordered potential landscape and the spatially dependent, open-dissipative terms of the rotating pump. Tracking the average distance of this dynamic phase defect as a function of frequency (Fig. 4F) reveals strong agreement with the experimental results, with vortices (antivortices) approaching the center as the rotation is increased for positive (negative) frequency. In particular, the decrease in the size of the error bars at $|\Delta f| \sim 5$ GHz is indicative of a transition from freely moving vortices to a captured vortex. The clear trends in this figure demonstrate that the main effect driving the accumulation of angular momentum, even in the theory, is not the stochastic formation of vortices driven by random initial conditions, but rather the dynamical nucleation and capture of vortices.

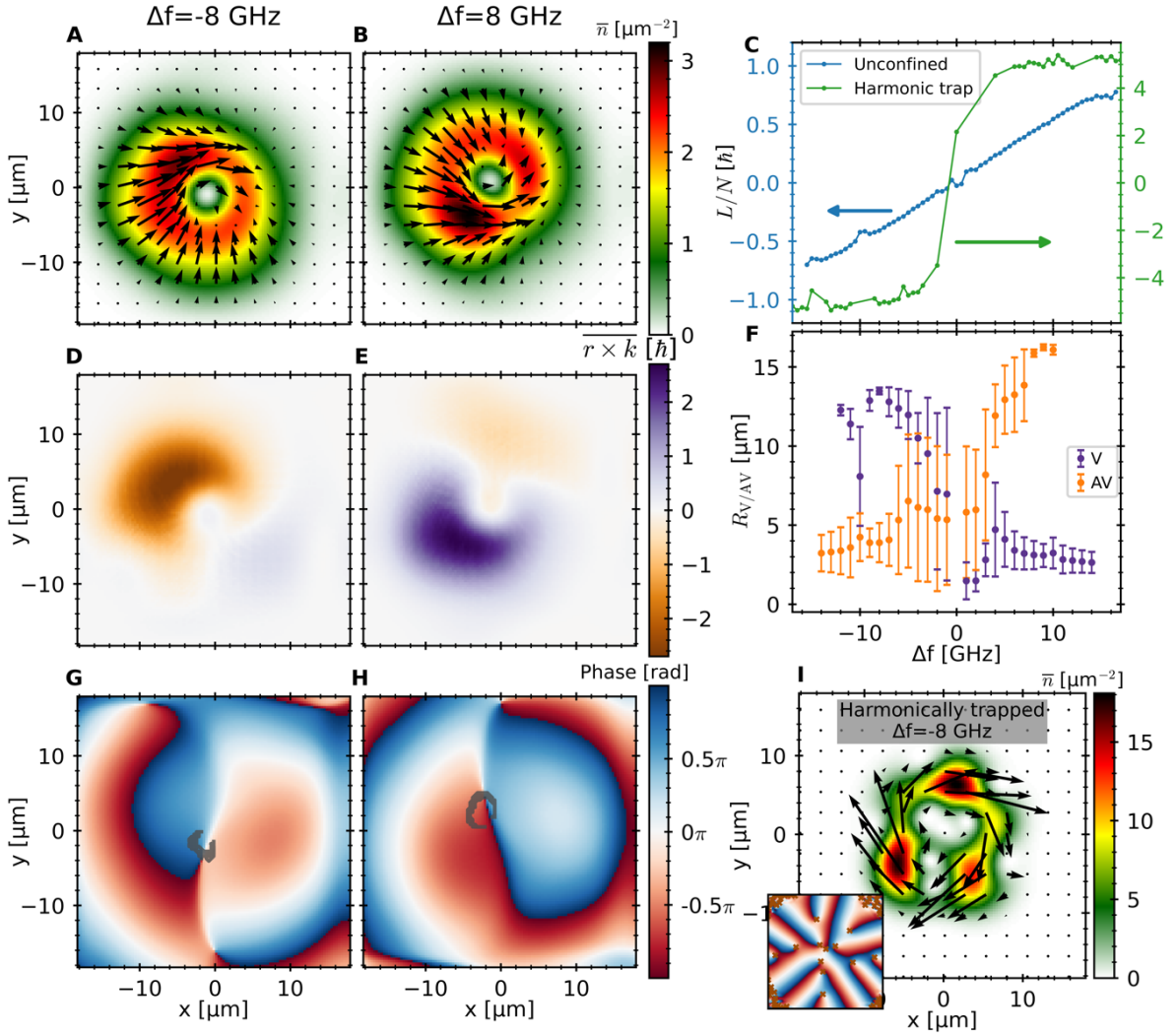

*Figure 4: Numerical simulations of rotating condensate dynamics. (A,B) Time-averaged condensate density and vector flow map at two different rotation frequencies. (C) Total angular momentum per particle $l = L/N$ as a function of pump rotation frequency for unconfined and harmonically confined condensates (see Methods). (D,E) Momentum distribution of the velocity maps in (A,B). (F) Vortex (orange) and antivortex (purple) distance to axis of rotation as a function of frequency. Error bars are the standard deviation in the vortex position over the last 2 rotation cycles in the simulation (G,H) Instantaneous phase profile for the distributions in (A,B). Grey regions indicate the position of the limit-cycle motion of the vortex core. (I) Instantaneous condensate density and vector flow map under the same parameters as (A) with a confining harmonic potential additionally applied. Inset shows the condensate phase with antivortex positions in dark orange crosses.*

Within the theoretical framework studied here, the microscopic origin for the transfer of pump angular momentum to the polariton condensate arises from a combination of two potential energy terms. The first is mechanical stirring provided by the repulsive interaction with a rotating pump-induced excitonic reservoir, which creates a real-valued potential in analogy with the trap-deformation method of stirring atomic condensates (*5*, *6*). The second is the gain-loss dynamics of the same rotating reservoir – i.e. an imaginary-valued potential. The relative contributions of these two stirring mechanisms are likely to vary with experiment and sample parameters but for the simulations in Figure 4 both terms were found necessary to describe the rotation incorporated in our experiment (see Supplementary 11).

Finally, the radial expansion of the unconfined, inhomogeneous condensate (Fig. 2A-C) means that vortex capture is partially inhibited by the presence of Magnus forces (*56*) which push vortices azimuthally, preventing capture in high density regions (*57*). The addition of an external potential can suppress these forces leading to a much larger number of nucleated vortices under otherwise equivalent pumping conditions. To confirm this, we add a harmonic trapping potential ($\omega \sim 505$ GHz, an order of magnitude larger energy than the disorder) to the simulations. The angular momentum per particle is now $\sim 10 \times$ that of the identical parameters without a trapping potential in Fig 4A ($L/N \sim 5\hbar$ at high frequency). At a pump rotation frequency of $\Delta f = -8$ GHz, Fig 4I shows that numerous quantized vortices are nucleated, stabilized, and develop some degree of order. The total angular momentum of the harmonically confined condensate saturates considerably faster at larger $|\Delta f|$ than for the unconfined condensate (Fig 4C). Unlike for the unconfined condensate, the origin of this saturation is dominated by the energetic unfavourability of adding additional vortices to a given condensate area. At an average condensate density of $n_c \sim 10$ μm$^{-2}$ the estimated single vortex core area $\pi \xi^2 \approx 50$ μm$^2$ (where $\xi$ is the approximate vortex size - see Methods and Supplementary 12), is already a significant fraction of the $\approx 250$ μm$^2$ condensate area.

## Discussion

We have reported the continuous, off-resonant injection of angular momentum into a spontaneously formed exciton-polariton condensate, using a rapidly rotating optical pump. Evidence for driven rotation is obtained by spatially mapping the condensate flow and characterizing the angular momentum, finding that the condensate responds to the rotating pump up to frequencies of $\pm 7$ GHz with flow direction controlled by the handedness of the pump. Angular momentum is measured up to $\sim 1.5\hbar$/particle, sufficiently exceeding the critical angular velocity necessary for nucleation of quantized vortices. Correspondingly, we observed the deterministic nucleation and incorporation of quantized vortices with a handedness consistent with pump rotation direction, providing independent verification that we have accurately quantified the condensate angular momentum. This quantity of angular momentum is understood to be limited in the present experiment by the lack of a deep, external confining potential, with simulations showing that this technique is capable of injecting considerably larger amounts of angular momentum and nucleating numerous vortices if a suitable trapping landscape is incorporated. Our demonstration of a scheme to generate a fast-rotating condensate, capable of nucleating multiple and dynamic vortices without coherent imprinting in analogy to the phenomenology observed in rotating atomic BEC, opens up numerous opportunities for study of superfluidity and quantized vortex ordering in an open-dissipative condensate and will lead to new ways of generating topological photonic matter in a highly non-linear and controllably non-Hermitian platform.

## Methods

*Semiconductor microcavity sample* – The microcavity sample used consists of 27(23) AlAs/Al$_{0.2}$Ga$_{0.8}$As mirror pairs for the bottom(top) distributed Bragg reflectors (DBRs) forming a $\lambda/2$ cavity with 3 sets of 4 × 13 nm wide GaAs/AlAs QWs, one set placed at the anti-node of the cavity and one set in each of the two first mirror pairs. The Rabi splitting is $2\hbar\Omega \approx 9.2$ meV.

*Lasers* – Two low-noise lasers are used: a cavity-stabilized Spectra-Physics Matisse CW Ti:Sapphire (time-averaged linewidth $< 200$ kHz) and a Toptica TA Pro diode laser-pumped tapered amplifier (typical linewidth $\sim 50$ kHz). A small fraction (<5%) of each beam is picked off, combined, and filtered through a single-mode fiber, with resulting temporal interference measured on a fast photodiode (Hamamatsu G4176-03, 30 ps response time). The beat signal is used to monitor the frequency offset of the Toptica from the Matisse ($|\Delta f| < 8$ GHz, limited by the electronic circuit bandwidth). Frequency tuning is achieved via a manually tuned piezo voltage that controls the diode's external cavity grating. A simple fork pattern on a Hamamatsu X-10468 spatial light modulator converts the Matisse laser into a LG mode of tunable charge. The two lasers are merged with a beam-splitter and put through a 20x objective to pump the sample. A LG orbital charge of $l = 2$ was chosen to create a reflection-symmetric modulation around the center whilst minimizing diffraction losses at the spatial-light modulator. The resulting superposition mode consists of $l = 2$ holes distributed on a ring with a radius determined by the relative amplitude of the two constituent modes. Both lasers are tuned to the first Bragg minimum of the microcavity ($\lambda_{pump} \sim 760$ nm). The spectral width of the first Bragg minimum is $\Delta\lambda \sim 0.85$ nm, significantly broader than the maximum used 20 GHz scanning range ($\sim 0.02$ nm) of the Toptica.

*Experimental apparatus* – Photoluminescence imaging measurements are conducted in a continuous-flow liquid helium cryostat, maintaining a fixed sample temperature of 4 K. An acousto-optic modulator with a 50 μs shutter speed is used to create quasi-CW, single-shot condensates while limiting sample heating. We utilize two sets of confocal optics to allow access to both the spatial and momentum coordinate Fourier planes, in addition to giving independent control over the resolution and field of view in both coordinate spaces. Light is imaged on a cooled CCD camera. Figure 5 shows a simplified schematic of the imaging optics.

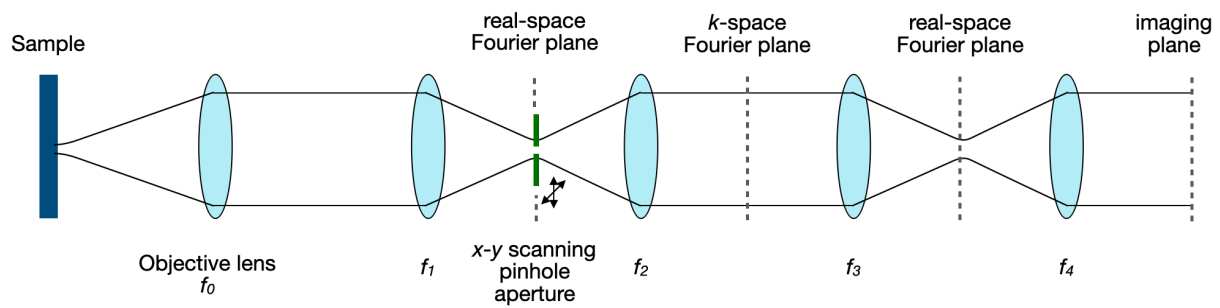

*Figure 5: Simplified schematic of confocal imaging optics setup, highlighting the positions of Fourier space planes and the scanning pin-hole aperture in the first real-space plane.*

To resolve the polariton momentum vector, a pinhole aperture is scanned in the spatial Fourier plane of the confocal optics setup. The average momentum at that pinhole position is then taken to be the centroid of the photoluminescence cloud in momentum space. The angular momentum is calculated using $\langle L \rangle = \int dr\, \rho(r)\, (r \times v(r)) \approx \sum_r r \times k_{centroid}$, where $\rho$ is the mass density, $\boldsymbol{r}$ the position in the $(x,y)$ plane, $\boldsymbol{k}$ the polariton linear momentum and $\boldsymbol{k}_{centroid}$ the average momentum of the pinholed cloud. The pinhole diameter of 100 μm (~1.6 μm on the sample) is chosen to optimize the light throughput and balance the resolutions in the conjugate spatial and momentum coordinates. Separation between each pinhole is ~1.4 μm, and care is taken to calibrate the geometric overlap between adjacent positions to avoid over-estimating the angular momentum (see Supplementary 3). Interferometric phase measurements were performed using an equal time Michelson interferometer with a retroreflector on one arm (see Supplementary 4). The interference fringes are presented and analyzed directly, allowing for analysis of dynamic vortex trajectories which present as a spatial variations of the interference visibility.

*Sample position* – To find a sample position that incorporated sufficient angular momentum to capture vortices, we moved the sample in increments of $\sim 10 - 50$ μm while keeping the rotation frequency fixed at $|\Delta f| = 3$ GHz. Small scans of Figure 2F were reconstructed at each position, and while all showed frequency-dependent flow, not all showed clear azimuthal flow. Around 3 out of 15 sample positions showed transverse flow all around the axis of rotation. For each of those, the frequency dependence was measured, and the most symmetric one of those three datasets is presented in the main text (see Supplementary 7 for an additional one).

*Numerical simulations* - The two-dimensional condensate dynamics are numerically simulated using the open-dissipative Gross-Pitaevskii equation (ODGPE) (8), forms of which have been used extensively in studies of exciton-polariton excitations and dynamics (9). A set of two coupled equations, Eqs. (1)-(2) describe the time-dependent dynamics of a condensed bosonic state $\psi(\boldsymbol{r},t) = \sqrt{n_C(\boldsymbol{r},t)}e^{i\phi(\boldsymbol{r},t)}$ of lower polaritons (LP) (eq. 1) coupled to a reservoir of thermal polaritons $n_R(\boldsymbol{r},t)$ (eq. 2):

$$i\hbar \frac{\partial \psi(\mathbf{r},t)}{\partial t} = \left( -\frac{\hbar^2 \nabla^2}{2M} + V_E(\mathbf{r}) - \frac{i\hbar}{2}[\gamma_C - R n_R(\mathbf{r},t)] + g_C |\psi(\mathbf{r},t)|^2 \right.$$
$$\left. + g_R n_R(\mathbf{r},t) \right) \psi(\mathbf{r},t) \qquad (1)$$

$$\frac{\partial n_R(\mathbf{r},t)}{\partial t} = P_l(\mathbf{r},t) - \gamma_R n_R(\mathbf{r},t) - R n_R(\mathbf{r},t)|\psi(\mathbf{r},t)|^2 \qquad (2)$$

where $M$ is the polariton mass, $V_E$ is the external potential, $\gamma_C$ the photon decay, $R$ the scattering rate from the reservoir, $g_C$ the polariton-polariton interaction constant, $g_R$ the polariton-reservoir interaction constant, and $\gamma_R$ the reservoir decay.

The rotating laser pump $P_l$ is composed of a Laguerre-Gaussian and a Gaussian beams that are frequency offset:

$$P_l(\mathbf{r},t) = |A_1 \psi_1(\mathbf{r},t) + A_2 \psi_2(\mathbf{r},t)|^2$$

where

$$\psi_1(r,t) = \left(\frac{r\sqrt{2}}{w_1}\right)^{|l|} \exp\left(-\frac{r^2}{w_1^2}\right) L_p^{|l|}\left(\frac{2r^2}{w_1^2}\right) \exp(-il\phi)\exp(i\omega_1 t)$$

is a Laguerre-Gaussian with $l = 2$, with $L_p$ being the generalized Laguerre polynomial, and

$$\psi_2(r,t) = \exp\left(-\frac{r^2}{w_2^2}\right) \exp(i\omega_2 t)$$

is a Gaussian beam. The beam waists at the sample plane are $w_1 = 8$ μm and $w_2 = 8.5$ μm. The pump rotation frequency $\Delta f = (\omega_1 - \omega_2)/2\pi$ is set within the range $|\Delta f| \leq 20$ GHz, which drives the time-dependent relative phase $\Delta\theta(t) = t\Delta f$ between the two beams. The superposition amplitude is calibrated to the laser pump power.

Fixing the polariton mass and lifetime, and condensate size with those extracted from experiments, the polariton non-linearity, reservoir lifetime and scattering rate are varied for a qualitive match to experimental measurements. Parameters used in Fig. 4 simulations are LP effective mass $M \sim 4 \times 10^{-5} m_e^0$ ($m_e^0$ is the free electron mass), LP decay rate $\gamma_C \sim 0.1 \text{ ps}^{-1}$, reservoir decay rate $\gamma_R \sim 0.15 \text{ ps}^{-1}$, stimulated scattering rate $R \sim 0.03 \text{ μm}^2 \cdot \text{ps}^{-1}$, condensate self-interaction $g_C \sim 6 \text{ μeV} \cdot \text{μm}^2$ and condensate-reservoir interaction $g_C = 2 g_R$.

The external potential $V_E(\mathbf{r})$ is a randomly disordered landscape (see Supplementary 9) in all except for the data in Fig. 4e (green line and Fig. 4h, where a harmonic trap ($\omega \sim 505$ GHz) is used. Similar to the experiment, the simulations were carried out over several random disorder realizations, with the one presented having the closest match to experimental data. The condensed mode is seeded by a Gaussian distribution of similar size to the pump mode, with infinitesimal amplitude, and a random initial phase. This set of equations is solved using the split-step operator method.

Note added: While under review, we became aware of a related work using a distinct pumping scheme for off-resonantly driving rotation (*31*).

# Acknowledgements

The authors acknowledge financial support by the Japan Society for the Promotion of Science Grants-in-Aid for Scientific Research (JSPS KAKENHI) Grant Numbers JP17H04851 and JP19H0561, the Japan Science and Technology Agency (JST) PRESTO Grant Number JPMJPR1768, NTT Research and the State of Bavaria.

## Author contributions

M.D.F. conceived the project and carried out the numerical simulations. Y.dVI.R. took the experimental data. M.D.F and Y.dVI.R. constructed the experimental apparatus, analyzed and interpreted the data, and wrote the paper. C.S., S.K. and S.H. prepared the samples. All authors contributed to discussions and to the final manuscript.

## Competing Interests

The authors declare that they have no competing interests.

## Data availability

All data needed to evaluate the conclusions in the paper are present in the paper and/or the Supplementary Materials.

# Supplementary Information

**Experiments**

1. Condensate density estimation

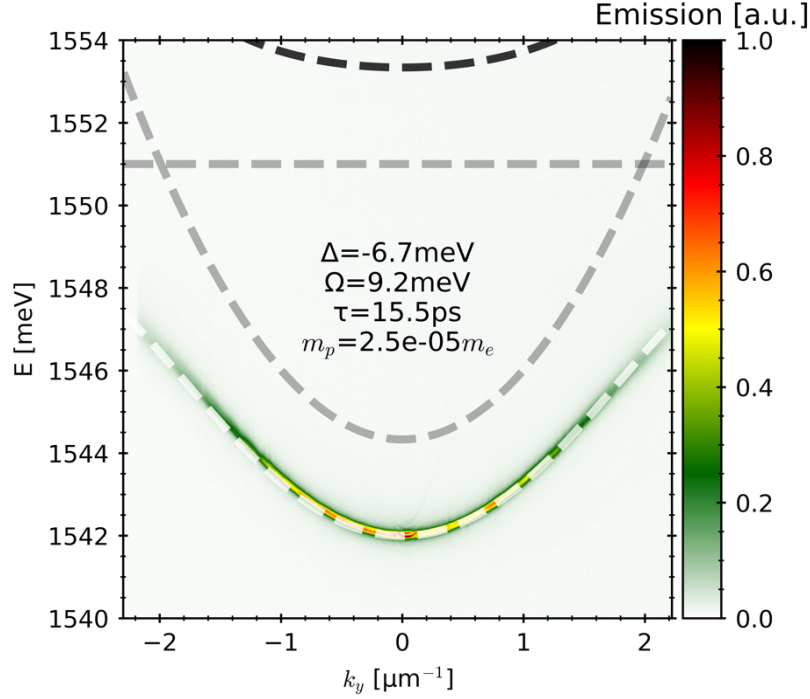

***Figure S1: Polariton dispersion.*** *Low power angular-resolved emission of the lower polariton. Dashed lines are the least-square fit of a 2-mode photon-exciton coupling Hamiltonian: exciton and photon dispersions in grey, upper polariton in black, lower polariton in white.*

Following the approach in (*58*), the condensate density is estimated from the CCD counts ($n_{counts}$) by first estimating the photon flux ($\Phi$) exiting the microcavity,

$$\Phi = \frac{\alpha\, n_{counts}}{\eta\, \tau_{exposure}},$$

where $\alpha = 4e^-/count$ is the photoelectron sensitivity and $\eta = 0.054$ is the total detection efficiency taking into account both the camera's quantum efficiency and the transmission through the optical path. Finally, the condensate number density $n_c$ is estimated by equation

$$n_c = \frac{\Phi\, \tau_c}{|C|^2},$$

where $\tau_c \sim 15.5$ ps and $|X|^2 = 0.21$ are directly measured from the low power dispersion of the microcavity (Fig. S1).

2. Condensation threshold when pumped by both lasers

In the presence of both Gaussian and $l = 2$ Laguerre-Gaussian pumps, the condensate transition is evidenced by an abrupt transition in the integrated output photoluminescence intensity at a threshold power of $P_{th} \sim 70$ mW (see Fig. S2). The condensate density then has an approximately superlinear dependence on power. Experimentally, the power above threshold can be varied within the range $\bar{P} = P_{laser}/P_{th} \sim 1 \rightarrow 2.57$.

The linewidth of the emission also reduces at threshold from ~200 µeV to ~70 µeV, which is limited by the resolution of our spectrometer. These linewidths are larger than all rotational energies in the experiment (10 GHz ≈ 40 µeV).

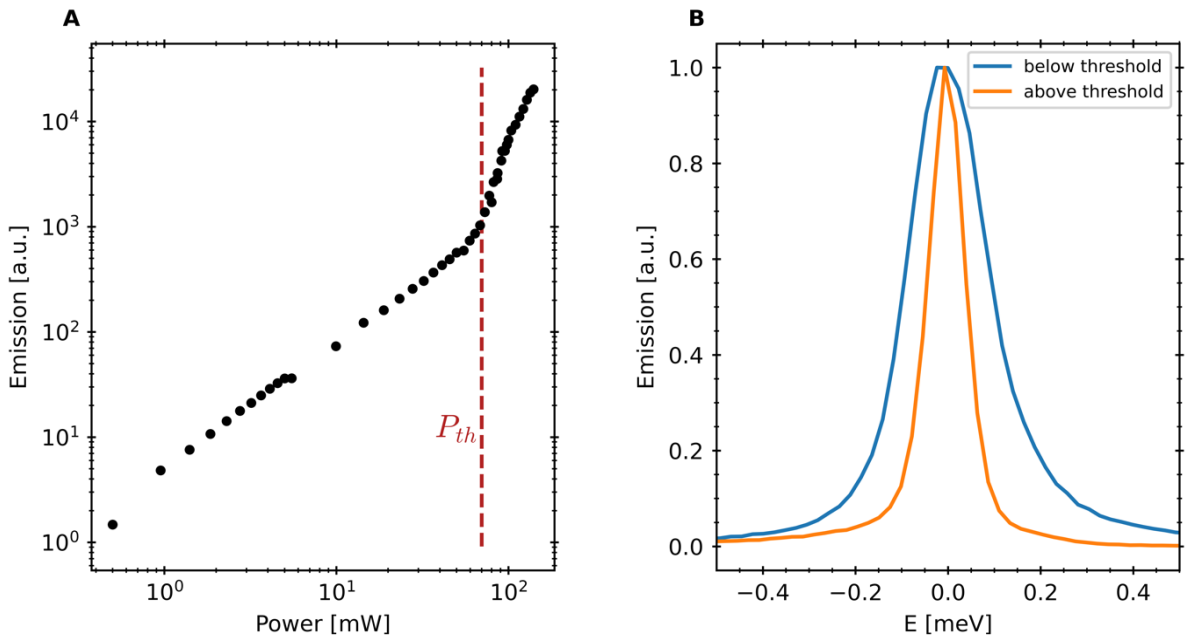

*Figure S2: Condensation transition and linewidth. (A) Integrated photoluminescence intensity as a function of total laser power. The threshold $P_{th} \sim 70$ mW is indicated by the red vertical line. (B) Linewidth at $k\sim0$ below and above threshold*

3. Pinhole overlap in angular momentum measurement

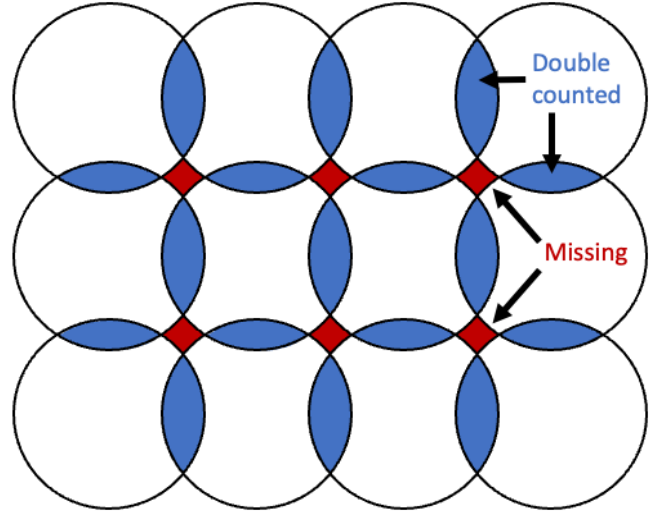

***Figure S3: Diagram of pinhole overlap.*** *Black circles indicate the edge of the pinhole at each measurement position. Blue are areas that are double counted in the density and angular momentum measurement, red are areas that are not measured.*

Each measurement of momentum (Fig. 2c) gives the linear momentum of a section of the condensate of $\sim 2.6$ µm². When scanning the pinhole across the condensate in a square grid (Fig. S3), sections of the condensate get pinholed more than once (blue), while other sections are not measured (red). We assume the overall momentum of the condensate changes smoothly in between measurements, when calculating the total angular momentum $L = \sum_{pinholes} r \times k$, we add the following correction:

$$A_{intersection} = 2 R^2 \cos^{-1}\frac{d}{2R} - \frac{d}{2}\sqrt{4 R^2 - d^2}$$
$$A_{missing} = d^2 - \pi R^2 + 2 A_{intersection}$$
$$L_{correction} = A_{missing} \sum_{ijkl} \frac{L_i + L_j + L_k + L_l}{2} - A_{intersection} \sum_{ij} \frac{L_i + L_j}{2}$$

Where $R = 1.6$ µm is the pinhole radius, $d = 1.4$ µm is the distance between pinhole measurements, and the sums are carried over $ij$ nearest-neighbour pairs, and $ijkl$ nearest 2x2 tiles.

4. Experimental Interferometry

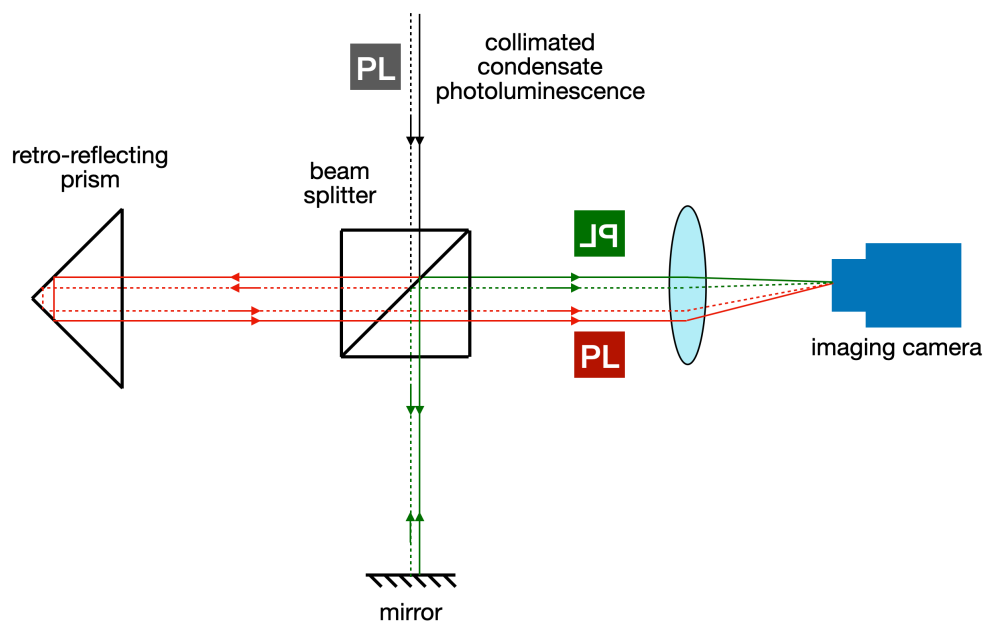

*Figure S4: Sketch of Michelson interferometer used for relative phase measurements. The condensate photoluminescence enters from the top as a collimated beam and is split into two arms, one of which is retro-reflected. The images are recombined on a camera with a controlled angle and spatial offset to generate a suitable interferogram, in addition to one of the arms being a spatial mirror image. The image orientation is indicated by the **PL** notations.*

The spatial phase coherence of the condensate was measured using a Michelson interferometer at zero-time delay, with a retroreflector in one of the arms (Fig. S4). The resulting interference pattern shows extended spatial coherence even for rotating condensates (Fig. S5a). Each arm image is spatially offset by ~5 μm (blue circles in Fig. S5b). Off-axis phase retrieval (*59*) is used to separate the spatial distribution of the intensity (Fig. S5b) and of the phase (Fig. S5d) from a single interference image. A single vortex is clearly observable by the presence of two forks in the normalised fringe image (Fig. S5c, red circles), with each fork corresponding to each of the two Michelson arms. The vortex can also be seen in the retrieved phase of the condensate (Fig. 5d).

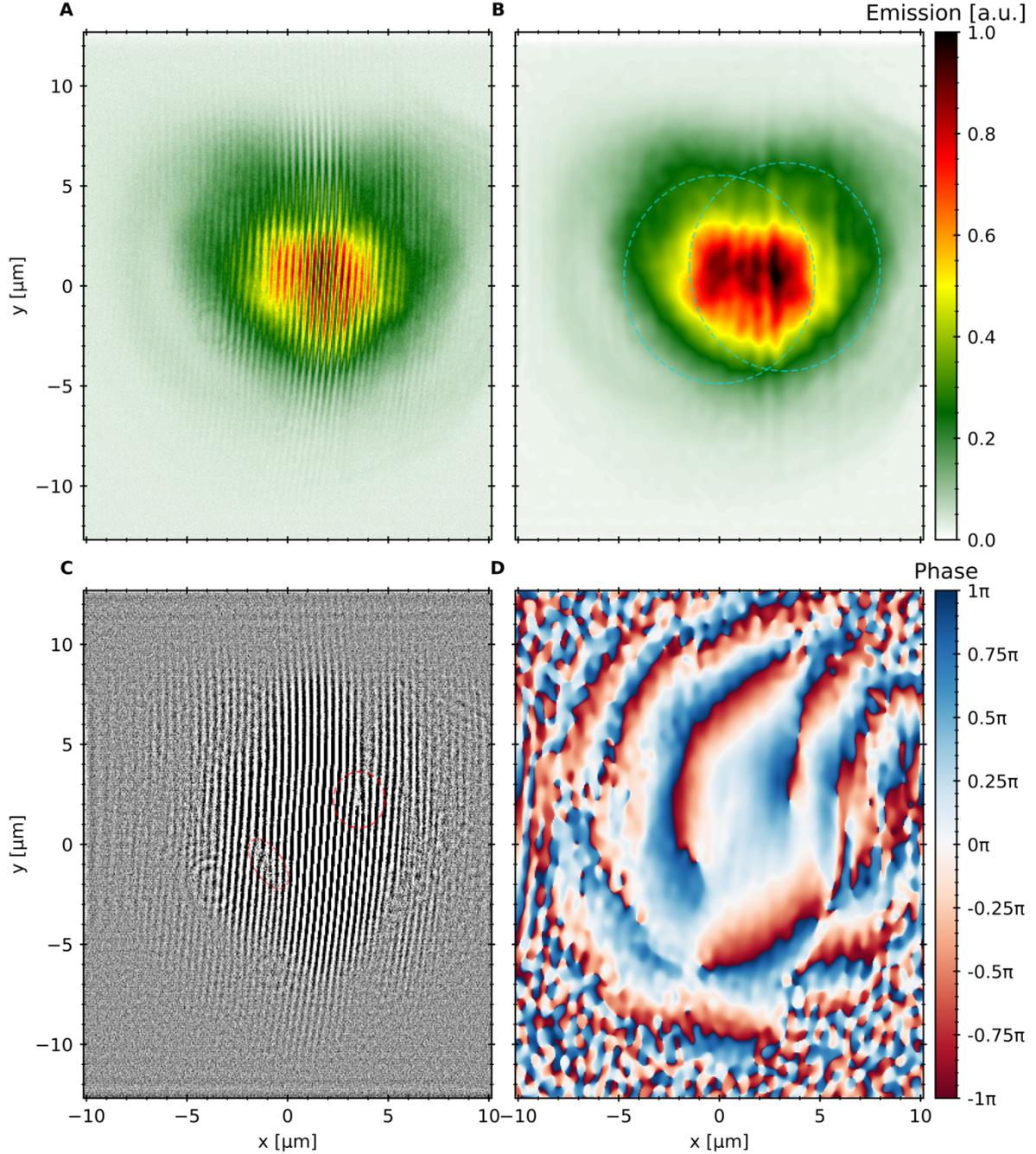

*Figure S5: Spatial coherence and single-quantized vortex. **A** Interference of condensate field with its retro-reflected image for $\Delta f = -6$ GHz. **B** Condensate emission without interference fringes. Blue circles indicate the approximate position of the images coming from each of the Michelson interferometer arms. **C** Fringes with the background density variation removed. Red circles indicate the position of the vortex. **D** Extracted phase map. B-D are reconstructed using off-axis phase retrieval.*

5. Raw data of shot-to-shot vortex positions

To study the shot-to-shot variations for different condensate realizations, for each pump rotation frequency in the range $-7 \leq \Delta f \leq 7$ GHz, 10 interferometric images are captured.

Both at high rotation speed (Figure S6) and low speed (Figure S7), every captured frame is qualitatively the same. There are shot-to-shot variations that lead to slightly different measured vortex positions, and ~10% realizations (#10 in S6, #6 in S7) have fringes that have been sufficiently blurred out that it is difficult to claim whether a vortex is still there or not. This variation is likely due to the stability of the Michelson interferometer used and does not take away from the conclusion that vortices are *deterministically nucleated and captured*.

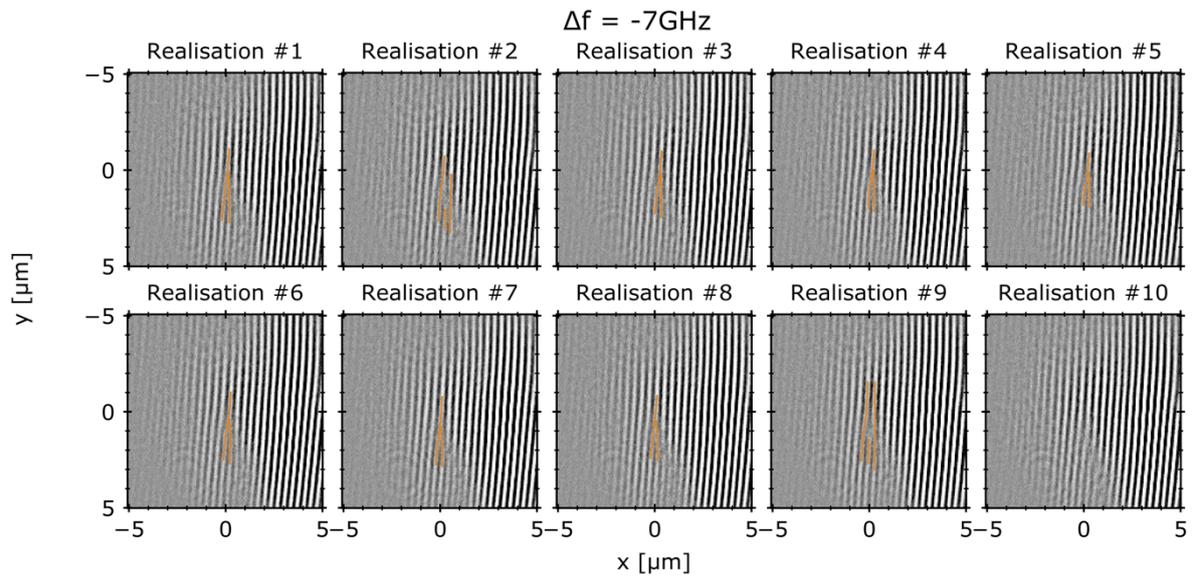

***Figure S6:*** *Single-shot interferograms at $\Delta f = -7\ GHz$. Each subpanel corresponds to an independent realization of the condensate.*

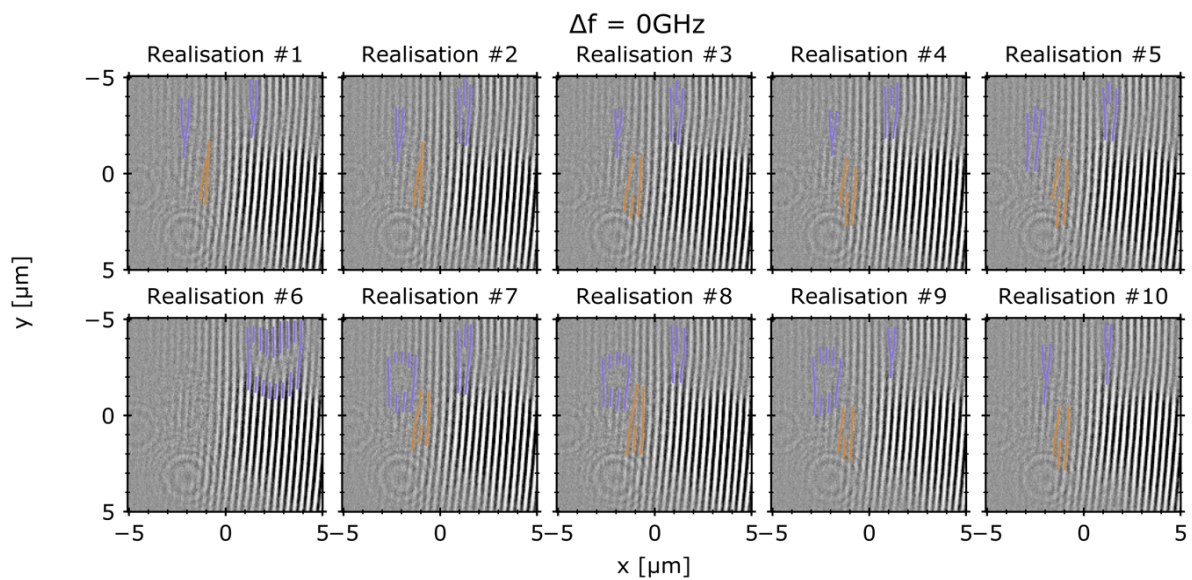

***Figure S7:*** *Single-shot interferograms at $\Delta f = 0\ GHz$. Each subpanel corresponds to an independent realization of the condensate.*

6. Raw data of vortex position as a function of frequency

Given 90% of condensate realizations result in qualitatively similar vortex positions, the positions of the vortices can be tracked as the frequency is varied by looking at a single characteristic frame for each frequency (Fig. S8). From these images the vortex positions in Fig. 3F are extracted.

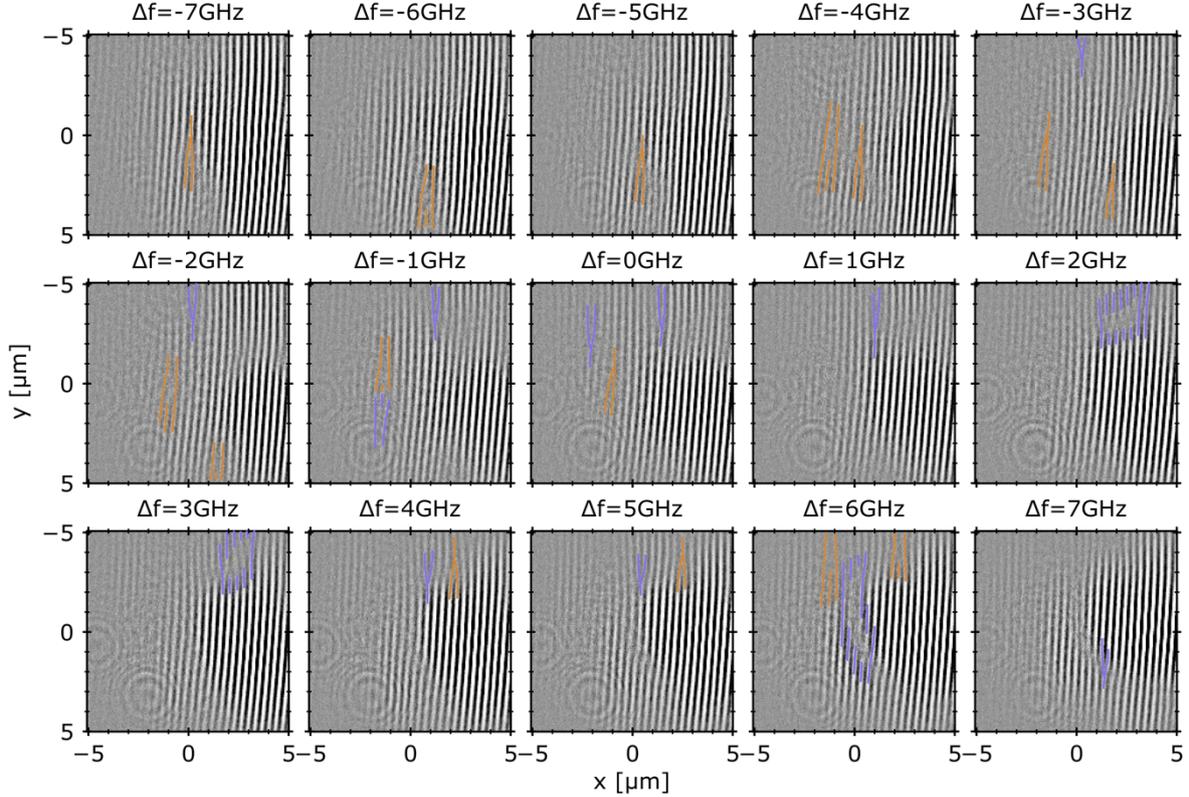

*Figure S8: Single-shot interferograms as a function of frequency.* *The purple (orange) lines are guides to the eye used to measure the vortex (antivortex) location and error.*

7. Measured angular momentum at a different sample position

Experiments frequently exhibit an asymmetric response with frequency, where the angular momentum per particle ($L/N$) has a different magnitude depending on rotation direction. This behaviour arises from the presence of microcavity sample disorder. Disorder is dominated by inhomogeneous quantum well broadening (*55*) and leads to a spatial variation in the energy of the ground state.

Given that disorder plays an important role, we studied multiple sample positions (see Methods) and presented the results for the sample position that demonstrated the most symmetric angular momentum acquisition. Figure S9 shows the same results as Figure 3 in the main text, but measured at a different sample position. The main qualitative features remain: azimuthal flow controlled by the frequency (A, B), an overall change in handedness when changing rotation direction (D, E) and a continuous incorporation of angular momentum (C). Interferometry was not performed at this sample position.

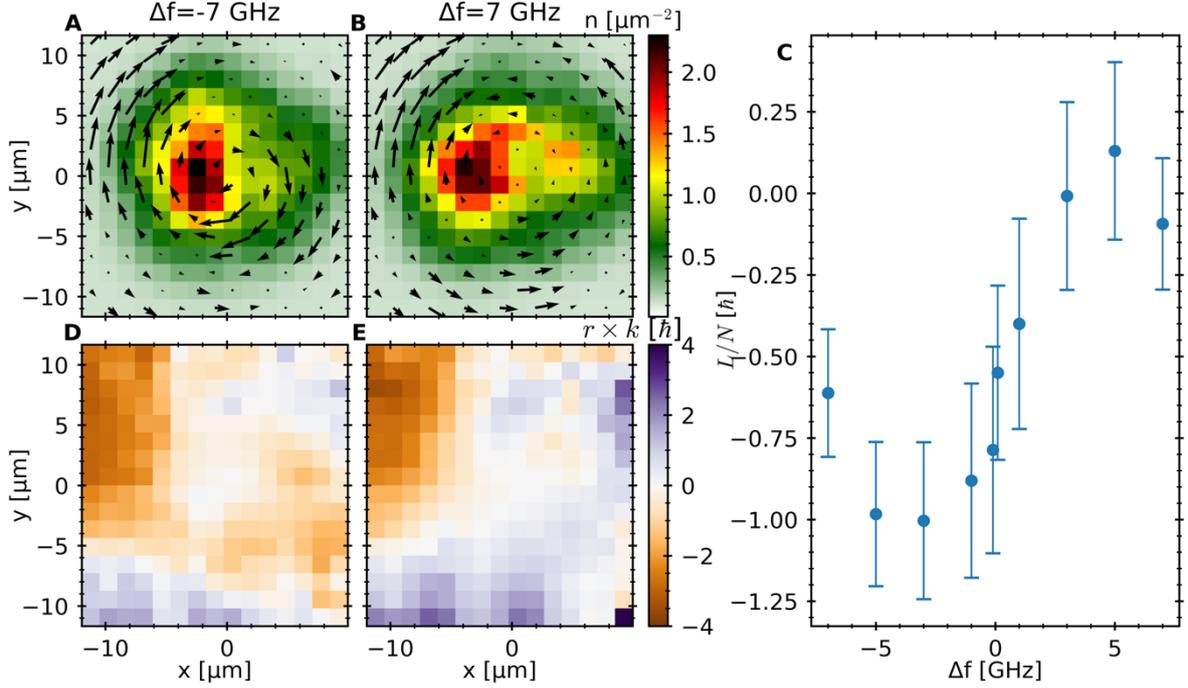

*Figure S9: Angular momentum acquisition at a different sample position. (A,B) Condensate density and azimuthal velocity map extracted from spatially pin-holed data at two different rotation frequencies. (C) Angular momentum per particle $l = L/N$ across the measurement area as a function of pump rotation frequency. Main source of error is the position of the axis of rotation, which is estimated to be $\pm 1.5$ µm. (D,E) Angular momentum distribution of the velocity maps in (A,B).*

8. Lower polariton effective mass

The effective mass is an important parameter for the study of quantized vortices. The vortex size is partially determined by the mass $M$ (through specification of the healing length parameter $\xi = \hbar/\sqrt{2Mg_c n_c}$, where $g_c n_c$ is the condensate nonlinearity).

Figure S10 shows a plot of the experimentally extracted polariton effective mass as a function of the exciton fraction $|X|^2$, where $X$ is the exciton Hopfield coefficient. The semiconductor microcavity samples embed a spatial variation in exciton-photon detuning with position, and thus the exciton fraction is varied by altering the measurement position on the sample.

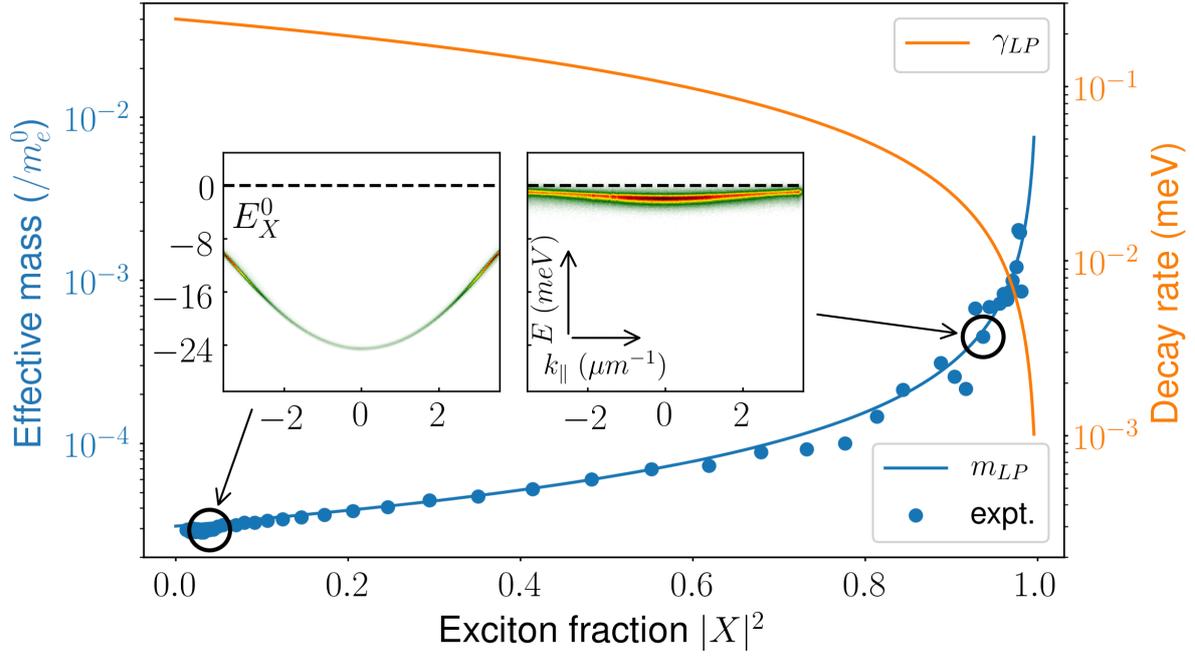

*Figure S10: Measured effective mass of lower polaritons.* Experimental polariton mass data points with theoretical fit, and an estimated corresponding decay rate, with variation in exciton fraction. Inset panels show examples of polariton dispersion at small and large mass as labelled.

The experimental data mass points are extracted from a parabolic fit to the measured lower polariton dispersion (see inset for two specific examples). The LP effective mass ($m_{LP}$) is fit by a weighted average of the exciton ($m_X$) and photon ($m_C$) masses.

$$\frac{1}{m_{LP}} = \frac{|X|^2}{m_X} + \frac{|1-X|^2}{m_C}$$

While the LP decay rate ($\gamma_{LP}$) is not measured, the estimated curve is shown which is also determined via the exciton ($\gamma_X$) and photon ($\gamma_C$) decay rates.

$$\gamma_{LP} = |X|^2 \gamma_X + |1-X|^2 \gamma_C$$

Notably, this particular GaAs microcavity sample exhibits a LP mass tunability over approximately 2 orders of magnitude, thereby allowing the vortex size to also vary over a wide range.

**Numerical Simulations**

9. Disorder potential

   To numerically emulate the influence of experimental sample disorder, a shallow disordered potential landscape is added to the Gross-Pitaevskii simulations, and the results are shown in Fig. 4. The specific form is a randomly distributed placement of islands with a Gaussian distributed width and depth, with the parameters:
   
   Mean spatial width $\Delta x = 5$ μm  
   Spread of width $\sigma = 1.25$ μm  
   Mean depth $\Delta V = 100$ μeV

   An example of a typical disorder landscape is given in Fig S11.

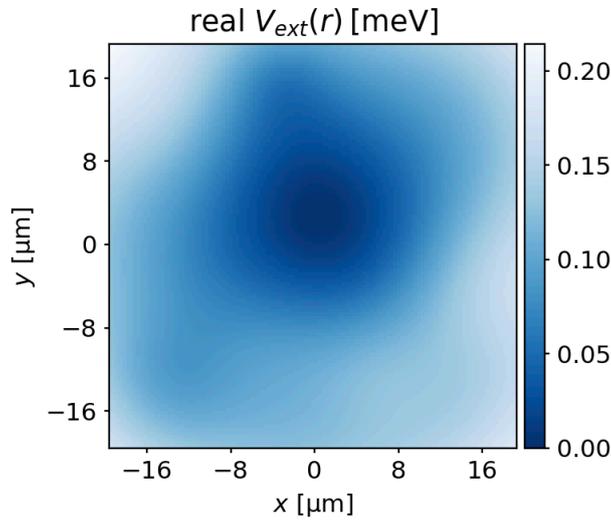

*Figure S11: Disorder profile used in the simulations of Figure 4.*

10. Observed angular momentum as a fraction of $\hbar$/particle

As discussed in the main text, there are several possible ways in which an angular momentum of $< 1\ \hbar$/particle can appear in our experiments and in our simulations. Here we introduce the analytical results known for off-axis vortices in cylindrically symmetric BECs, and expand on the idea of dynamic and transient vortices:

   a. **Off-axis vortices** are vortices present for the whole duration of the measurement, but not aligned with our estimated axis of rotation. Our toy model is a harmonically trapped condensate of radius $R_\perp$ in the Thomas-Fermi approximation (*53*). The angular momentum per particle for an ensemble of $N_v$ singly-quantized vortices at arbitrary positions $r_{\perp 0 i}$ is:

$$\langle l \rangle \approx \sum_{i=1}^{N_v} \hbar \kappa \left[ 1 - \left( \frac{r_{\perp 0 i}}{R_\perp} \right)^2 \right]^2$$

where $\kappa = \pm 1$ is the vortex sign.

For a single vortex, $\langle l \rangle$ can only be $1\hbar$ when the vortex is on the rotation axis ($r_{\perp 0} = 0$). For any other positions, it contributes less than $1\hbar$ to the total angular momentum. For an ensemble of $\kappa = \pm 1$ quantized vortices, $\langle l \rangle$ may take on an arbitrary and continuous value. Since only one vortex can be on-axis, a lattice of $N_v > 1$ vortices necessarily has an angular momentum $\langle l \rangle < N_v$ (60).

b. **Dynamic or transient vortices**. Each experimental snapshot is $\sim 10^6 \times$ longer than the polariton lifetime. Hence given the open-dissipative environment, it is possible for there to be vortices present for part, but not all of the measurement time. Extending from equation above, for an integration time of $T$, each dynamic vortex contributes an amount of angular momentum per particle of:

$$\langle l \rangle \approx \frac{1}{T} \sum_{i=1}^{N_v} \hbar \kappa \int_0^T \left[1 - \left(\frac{r_{\perp 0 i}(t)}{R_\perp}\right)^2\right]^2 dt$$

Again, this results in $\langle l \rangle$ being able to take an arbitrary and continuous dependency on the specific vortex trajectory and the average time it spends near axis of rotation of the condensate.

While the quantitative applicability of this model in the case of polariton condensates has not been tested, it does qualitatively demonstrate the physical underpinnings of the initially counterintuitive result that angular momentum is continuous.

11. Contributions of mechanical and non-Hermitian stirring

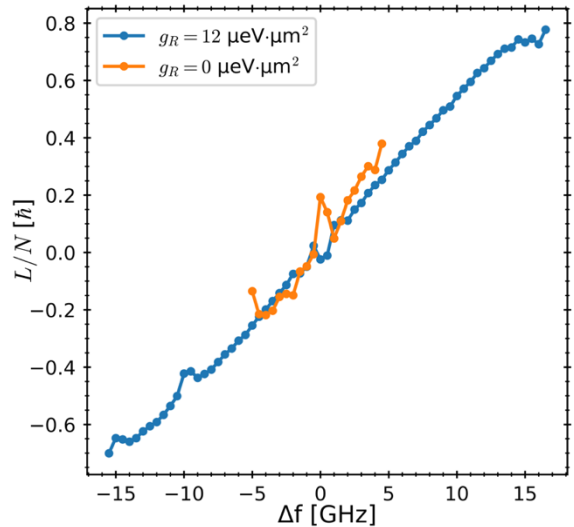

*Figure S12: Contributions of mechanical and non-Hermitian stirring. The numerically calculated total condensate angular momentum $L/N$ for finite reservoir-condensate interactions $g_R$ (both mechanical and non-Hermitian terms) and $g_R = 0$ (only non-Hermitian stirring).*

The relative contributions of mechanical and non-Hermitian stirring terms are estimated by numerical simulations for parameters in Figure 4. In these simulations, both the condensate-reservoir repulsive interaction ($g_R = 12$ µeV·µm$^2$) which drives mechanical stirring, and the gain-loss terms are incorporated (reproduced in Fig. S12, blue line). By setting $g_R = 0$, the non-Hermitian terms (i.e. spatially structured, time-dependent gain and loss) can be isolated (Fig. S12, orange line). We find that the contributions of both terms are crucial to fully capture the rotation dynamics, however the relative contributions of each depends heavily on the specific pump distribution and polariton simulation parameters. Nevertheless, for the experiments presented here, the best numerical correspondence shows that non-Hermitian terms dominate at small $|\Delta f|$, but fail to describe rotation at larger $|\Delta f| \gtrsim 5$ GHz, as the reservoir-condensate spatial overlap is reduced leading to loss of gain. This suggests that to rotate at the maximum possible velocities, both terms are important, and some improved design of the pump configuration to accommodate this may assist future experiments.

## 12. Estimated critical rotation frequency for vortex nucleation

Following (*45*), the critical rotation frequency necessary to nucleate a vortex in a rotating condensate is determined by:

$$\Omega_C = \frac{E_v}{L}$$

where the on-axis vortex energy for a uniform, finite-sized condensate (after background subtraction) is

$$E_v = \pi n \frac{\hbar^2}{M} \ln\left(1.464 \frac{R}{\xi}\right)$$

where $\xi = \hbar/\sqrt{2Mg_C n}$ is healing length, which determines the approximate vortex size, and the total angular momentum is $L = N\hbar$.

For polariton parameters in the ranges:

| | |
|---|---|
| Effective mass | $M \sim (2 \times 10^{-5} \to 10^{-3})m_e^0$ |
| Density | $n \sim 1 \to 100$ µm$^{-2}$ |
| Condensate radius | $R \sim 3 \to 15$ µm |

one can make the following estimates:

| | |
|---|---|
| Particle number | $N \sim 10^2 \to 10^5$ |
| Vortex radius | $\xi \sim 0.2 \to 10$ µm |
| Critical frequency range | $\Omega_C \sim 10^8 \to 10^{11}$ Hz |